# The vacuum and cryogenics system of the SOXS spectrograph

S. Scuderi[a], G. Bellassai[b], R. Di Benedetto[b], E. Martinetti[b], A. Micciché[b], G. Nicotra[c], G. Occhipinti[b], C. Sciré[b], M. Aliverti[d]*, M. Genoni[d], F. Vitali[e], S. Campana[d], R. Claudi[f], P. Schipani[g], A. Baruffolo[f], S. Ben-Ami[h], G. Capasso[g], R. Cosentino[i], F. D'Alessio[e], P. D'Avanzo[d], O. Hershko[h], H. Kuncarayakti[j], M. Landoni[d], M. Munari[b], G. Pignata[l], K. Radhakrishnan[f], A. Rubin[l], D. Young[m], J. Achren[n], J.A. Araiza-Duran[o], I. Arcavi[p], F. Battaini[f], A. Brucalassi[o], R. Bruch[h], E. Cappellaro[f], M. Colapietro[g], M. Della Valle[g], S. D'Orsi[g], A. Gal-Yam[h], M. Hernandez Diaz[i], J. Kotilainen[r], G. Li Causi[s], L. Marty[g], S. Mattila[j], M. Rappaport[h], D. Ricci[f], M. Riva[d], B. Salasnich[f], S. Smartt[m], R. Zanmar Sanchez[b], M. Stritzinger[t], H. Perez Ventura[i]

[a]INAF-Istituto di Astrofisica Spaziale e Fisica Cosmica di Milano, Via Corti 12, 20133 Milano Italy,
[b]INAF-Osservatorio Astrofisico di Catania – Via Santa Sofia, 78, 95123 Catania Italy,
[c]INAF – Istituto di Radioastronomia,
[d]INAF – Osservatorio Astronomico di Brera, Via Bianchi 46, I-23807 Merate (LC), Italy,
[e]INAF – Osservatorio Astronomico di Roma, Via Frascati 33, I-00078 Monte Porzio Catone, Italy,
[f]INAF – Osservatorio Astronomico di Padova, Vicolo dell'Osservatorio 5, I-35122 Padova, Italy,
[g]INAF – Osservatorio Astronomico di Capodimonte, Salita Moiariello 16, I-80131 Napoli, Italy,
[h]Weizmann Institute of Science, Herzl St 234, Rehovot, 7610001, Israel,
[i]INAF - Fundación Galileo Galilei, Rambla J.A. Fernández Pérez 7, E-38712 Breña Baja (TF), Spain,
[j]Tuorla Observatory, Department of Physics and Astronomy, University of Turku, FI-20014, Finland,
[k]Universidad Andres Bello, Avda. Republica 252, Santiago, Chile,
[l]ESO, Karl Schwarzschild Strasse 2, D-85748, Garching bei München, Germany,
[m]Queen's University Belfast, School of Mathematics and Physics, Belfast, BT7 1NN, UK,
[n]Incident Angle Oy, Capsiankatu 4 A 29, FI-20320 Turku, Finland,
[o]INAF-Osservatorio Astrofisico di Arcetri, Largo E. Fermi 5, I-50125, Firenze, Italy,
[p]Tel Aviv University, Israel,
[r]FINCA - Finnish Centre for Astronomy with ESO, FI-20014 University of Turku, Finland,
[s]INAF - Istituto di Astrofisica e Planetologia Spaziali, Via Fosso del Cavaliere, I-00133 Roma, Italy,
[t]Aarhus University, Ny Munkegade 120, D-8000 Aarhus, Denmark

## ABSTRACT

SOXS (Son Of X-Shooter) is a single object spectrograph built by an international consortium for the ESO NTT telescope. SOXS is based on the heritage of the X-Shooter at the ESO-VLT with two arms (UV-VIS and NIR) working in parallel, with a Resolution-Slit product ≈ 4500, capable of simultaneously observing over the entire band the complete spectral range from the U- to the H-band. SOXS will carry out rapid and long-term Target of Opportunity requests on a variety of astronomical objects.

The SOXS vacuum and cryogenic control system has been designed to evacuate, cool down and maintain the UV-VIS detector and the entire NIR spectrograph to their operating temperatures. The design chosen allows the two arms to be operated independently. This paper describes the final design of the cryo-vacuum control system, its functionalities and the tests performed in the integration laboratories.

*Contact author: salvatore.scuderi@inaf.it

# 1. INTRODUCTION

SOXS[1] (Son Of X-Shooter) is a single object spectrograph built by an international consortium for the ESO NTT telescope.

To accomplish this task SOXS, that is based on the heritage of the X-Shooter at the ESO-VLT, has two arms, the UV-VIS [2] and NIR spectrograph [3] working in parallel, that need to be operated at cryogenic temperatures. In particular, in the case of the UV-VIS spectrograph the CCD detector has to operated at 173 K while in the case of the NIR spectrograph the optical bench and the optics are kept to 150 K while the detector has an operating temperature of 40 K.

The vacuum and cryogenic control system has then been designed to evacuate, cool down and maintain the vacuum state of the vessels and the values of temperature within the required range.

# 2. SOXS VACUUM SYSTEM

The architecture of the SOXS vacuum system is shown in Figure 1. The NIR vessel is custom made while the UV-VIS vessel is composed by a Continuous Flow Cryostat (CFC) provided by ESO [4] and a custom part [2], [5]. The two vessels are evacuated via an external pumping station that combines a turbo molecular pump (2) and pre-vacuum pump (1). The evacuation of the vessels can be done simultaneously or independently thanks to the (4a) and (4b) electro-valves. The total volume to be evacuated is of the order of 130 liters while the exposed surface is about 6.5 $m^2$.

For the two vessels, the interfaces to the external world are the electro-valves (4a) and (4b). A full range pressure gauge (3) is installed to monitor the pressure in the two vacuum lines, also acting as a pressure switch to provide the conditional information to start/stop the pump(s). Pumping station and insulating valves (4a) and (4b) are connected through DN40 flexible bellows.

Both vessels are equipped with a wide range vacuum gauge (5) in order to monitor permanently the residual vacuum pressure. In the case of NIR vessel there is a second vacuum gauge for redundancy.

The NIR vessel is equipped with a safety overpressure valve (8) to prevent damages in case of sudden increase of the pressure, and, for controlled re-pressurization, with a venting valve (10) and a gas connector (11) as interface to a nitrogen gas supply.

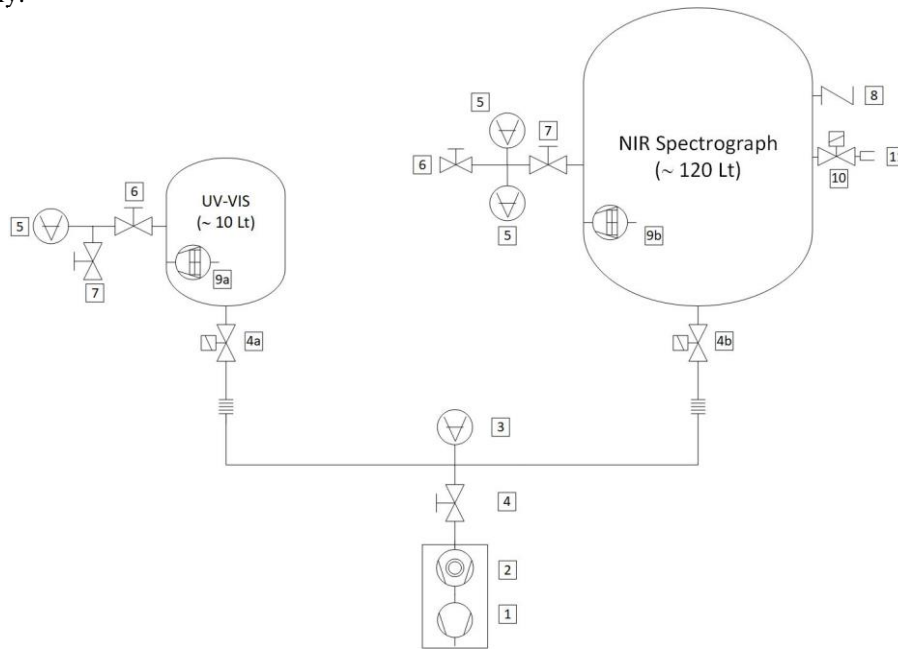

Figure 1. SOXS vacuum system architecture. See the text for the numbers.

Finally, both vessels are equipped with a cryo-pump (9) to guarantee the level of vacuum necessary for the operation of the two detectors.

The vacuum system (pre-vacuum pump, turbomolecular pump, valve (4) and sensor (3)) are not always online but they are disconnected by the SOXS spectrograph once the system is cold and the two electrovalves (4a) and (4b) are closed.

The vacuum components have been chosen according document ESO-046147 and are listed in Table 1.

Table 1. SOXS vacuum components

| Nr | Acronym | Component | Product | Supplier |
|---|---|---|---|---|
| 1 | PVP | Pre-vacuum pump | ACP 28 | Pfeiffer vacuum |
| 2 | TMP | Turbo-molecular pump | ATH 500M | Pfeiffer vacuum |
| 3 | PSW | Pressure Switch | WRG-D NW25 | Edwards vacuum |
| 4 | PVV | Pre-Vacuum valve | 215 377 | Leybold |
| 4a | IV1 | Insulation valve | 215 064 V01 | Leybold |
| 4b | IV2 | Insulation valve | 215 124 V01 | Leybold |
| 5 | VAG | Vacuum Gauge | WRG-D NW25 | Edwards vacuum |
| 6 | GE1 | Gauge Exchange Valve 1 | 110VSM025 + 110RTS025-V | Pfeiffer Vacuum |
| 7 | GE2 | Gauge Exchange Valve 2 | 110VSM025 | Pfeiffer Vacuum |
| 8 | SOV | Safety Overpressure valve | 89039 | Leybold |
| 9a | CP1 | UV-VIS Cryo-Pump | 13600-910000-0020 | ESO |
| 9b | CP2 | NIR Cryo-Pump | Custom made | INAF |
| 10 | N2V | Re-pressurization valve | 215 024 | Leybold |
| 11 | N2C | Nitrogen gas connector | Series QC4 | Swagelock |

## 2.1 Cryopumps

Due to their role in the evacuation process the cryo-pumps are considered part of the vacuum system.

The cryo-pump for UV-VIS cryostat is the standard one used by ESO in the CFC [4].

The cryo-pump used in the NIR vessel is instead designed specifically for it. The main body (see Figure 2) of the cryo-pump is made in copper. The pump is kept insulated from the optical bench by three G10 legs. This solution allows also exposing the lower surface of the cryo-pump. The upper and lower surfaces are covered with a copper grid. Inside its volume 0.2 litres of active charcoal can be accommodated. The Cryo-pump is connected to the cold head of the NIR vessel cryocooler by a thermal strap. Inside the cryopump there is a copper plate used to regenerate the cryo-pump. On the plate a PT103, to monitor its temperature, and a heating resistor, to heat it up, are mounted. The plate is kept insulated from the body through a G10 holder to avoid heating the cryocooler cold head.

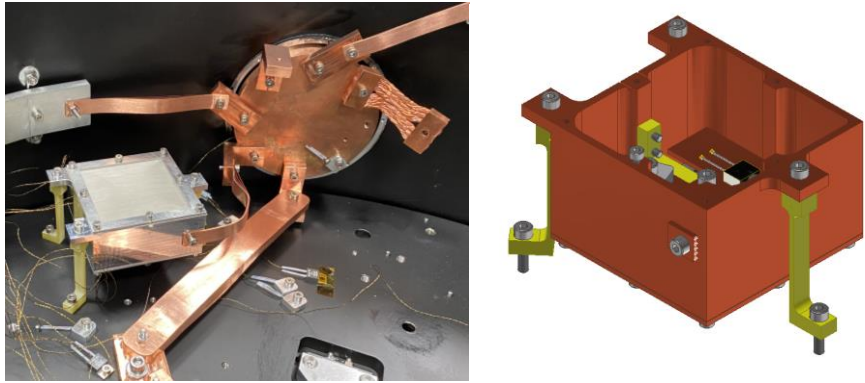

Figure 2. The cryo-pump for the NIR vessel. Left: external view of the cryo-pump showing also the copper strap to connect it to the cryocooler head. Right: internal view showing the copper plate insulated by the cryo-pump body with heater and PT100.

# 3. SOXS CRYOGENIC ARCHITECTURE

To allow the UV-VIS and the NIR spectrographs to be operated independently each one of them has its own cryogenic system.

## 3.1 UV-VIS spectrograph

In the case of the UV-VIS spectrograph the system used to cool down the CCD detector is a CFC. The CFC together with the detector head makes the UV-VIS cryostat (see Figure 3) as described in [2]. The UV-VIS cooling system includes also a LN2 storage tank and a vacuum insulated LN2 transfer line. The CFC operating principles are thoroughly described in [4] and will not repeated here.

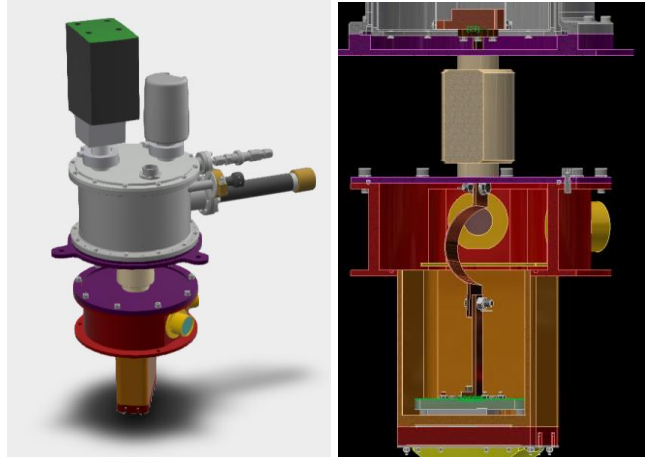

Figure 3. Left: mechanical design of the UV-VIS cryostat. Right: section of the UV-VIS cryostat showing the thermal connection between the CFC and the CCD.

The CFC is supplied by a portable Liquid Nitrogen storage tank of 120 litres capacity. Vacuum insulated lines are used for the transfer of LN2. These lines are custom made to the required length and interfaces. They connect the supply interface with the instrument interface via so-called Johnston couplings, furthermore, as SOXS is a rotating instrument, a special rotating feed-through for the LN2 supply of the cryostat is used [5]. The tank can be equipped with LN2 level sensors and the transfer line has a temperature sensor (PT100) to measure the temperature of the outer skin of the LN2 transfer line to raise a warning in case the vacuum inside the line deteriorates.

As stated in the introduction the only system that, within the UV-VIS spectrograph, has to be cooled down and thermally controlled is the CCD detector.

The CCD detector will be located in the "nose" of the UV-VIS cryostat (Figure 3). The thermal connection between CCD baseplate and the CFC cold finger is composed by three parts: a rod from the CFC cold finger to the detector holder, then a strap going inside the detector nose and again a rod to the CCD baseplate (Figure 3). This connection will allow to cool down the detector at 168K just below its operating temperature (173K). The CCD is mounted on an INVAR baseplate that will be connected to the UV-VIS cryostat window flange through four G10 spacers [2].

To control and monitor the temperature of the CCD detector a Lakeshore 336 temperature controller will be used. On the CCD baseplate a DT670-SD temperature sensor and a Caddock heating resistor MP820 will be mounted. The resistor will heat the CCD baseplate while the temperature sensor will give the feedback to the temperature controller.

## 3.2 NIR spectrograph

Figure 4 shows an external view of the NIR spectrograph. In this case, not only the detector but also the optical system (optics, mounts, bench), the thermal shield and the mechanisms will have to be cooled down.

The system selected to achieve this task is a closed cycle cooler (CCC), the 1-stage GM cooler COOLPOWER 250 MD coupled with COOLPACK 6000 HMD compressor both from Leybold. The cold head of this cryo-cooler has the advantage that can be operated in any orientation. This cryocooler has low vibration levels and the cold head is mounted on the vessel through a custom made anti-vibration mount. Another interesting feature is the possibility to vary the cold head motor speed allowing to adjust the refrigerating capacity of the head. Finally, the cryocooler can be remotely controlled.

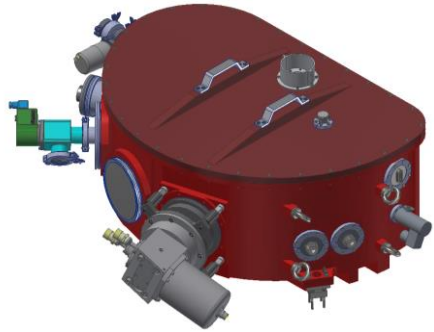

Figure 4. Design of the NIR spectrograph cryostat.

As stated in the introduction the entire spectrograph has to be cooled down at cryogenic temperatures. Apart for the cold head the coldest part in the vessel will be the detector whose operating temperature is 40 K to avoid persistence phenomena that can happen below 80K in HAWAII 2RG from Teledyne detectors. The required temperature stability for the detector is ±0.01 K. The optical bench and all the optics are instead operated at 150 K. Temperature stability is ±2 K for the optical bench and all the optics and mechanism apart from the prisms assembly that instead requires ±0.1 K.

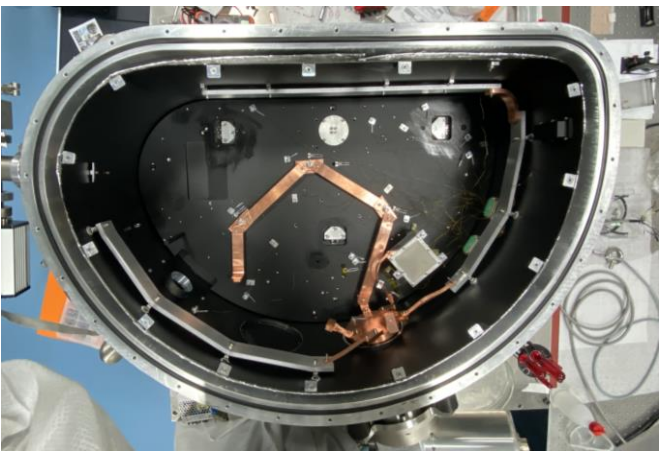
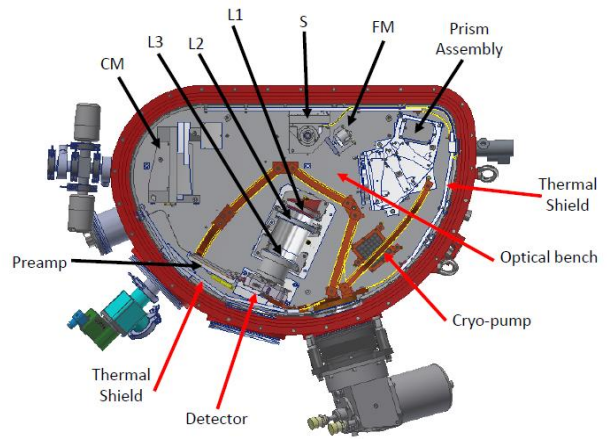

Figure 5. Left: internal view of the NIR cryostat showing the thermal connections of the optical bench to the CCC head. No optical elements are present. Right: drawing of the internal part of the NIR cryostat showing all the elements that have direct or indirect connection to the CCC head. For the letters see the text.

The following elements are directly connected to the CCC cold head through copper thermal straps:

1. NIR detector.
2. Optical bench.
3. Cryo-pump.
4. Thermal shield.

In the case of the NIR detector the connection to CCC is to its housing [3]. Like in the case of the CCD detector on the detector cold finger there is a DT670-CU temperature sensor and a Caddock heating resistor MP820. As security device against overtemperature a bimetal thermostat from Airpax is inserted in series to the resistor. The entire system is then redounded.

In the case of the optical bench on each of the three contact points a heating resistor is mounted close by while five temperature (PT103) sensors are placed in different places on the bench to give feedback.

In the case of the cryo-pump the copper thermal strap to the CCC is fixed on one of the external walls (see Figure 2). The temperature sensor and the heating resistor are mounted within its case as explained in section 2.1.

Finally, for the thermal shield two stripes starting from the CCC cold head will go left and right on the shield and a temperature sensor (PT103) will be mounted on the shield to monitor its temperature.

All the other optical elements will be connected not directly to the CCC but using dedicated thermal links to the optical bench (see Figure 5). In particular:

1. The pupil's stop and slit assembly (S).
2. The folding mirror (FM).
3. The collimator mirror (CM).
4. The prism assembly made by the collimator lens (CL), the three prisms (P1, P2, P3), and the grating (G).
5. The lens L1, L2, L3 and the filter (F) which together makes the camera assembly.
6. NIR detector preamplifier.

A heating resistor will be mounted on the optical element side of the thermal connection of the Camera Assembly, the Prisms Assembly and the Slit Assembly. Three temperature sensors (PT103) are placed on the Prisms Assembly, two on the Camera Assembly, one on the Slit Assembly and one on the NIR detector preamplifier.

## 4. SOXS CRYO-VACUUM CONTROL SYSTEM

The SOXS cryo-vacuum control system is based on a Siemens PLC Simatic S7-1500. The PLC attends to all the operations related to the vacuum and the cryogeny of the system except those strictly related to the detectors which are managed by the Lakeshore controller. Figure 6 shows the physical architecture of the SOXS cryo-vacuum control system with devices (pumps, valves, heating resistors) and sensors (pressure gauges and temperature sensors) managed through the Siemens PLC. The Lakeshore controller manages the active cool down and the warm up of the NIR detector and keeps UV-VIS and NIR detectors at their operating temperature once they are cold. The Lakeshore sends also alarms related to its functions to the Siemens. Finally, all cryo-vacuum parameters and settings and all the sensors readings are exposed to the external world through an OPC-UA interface.

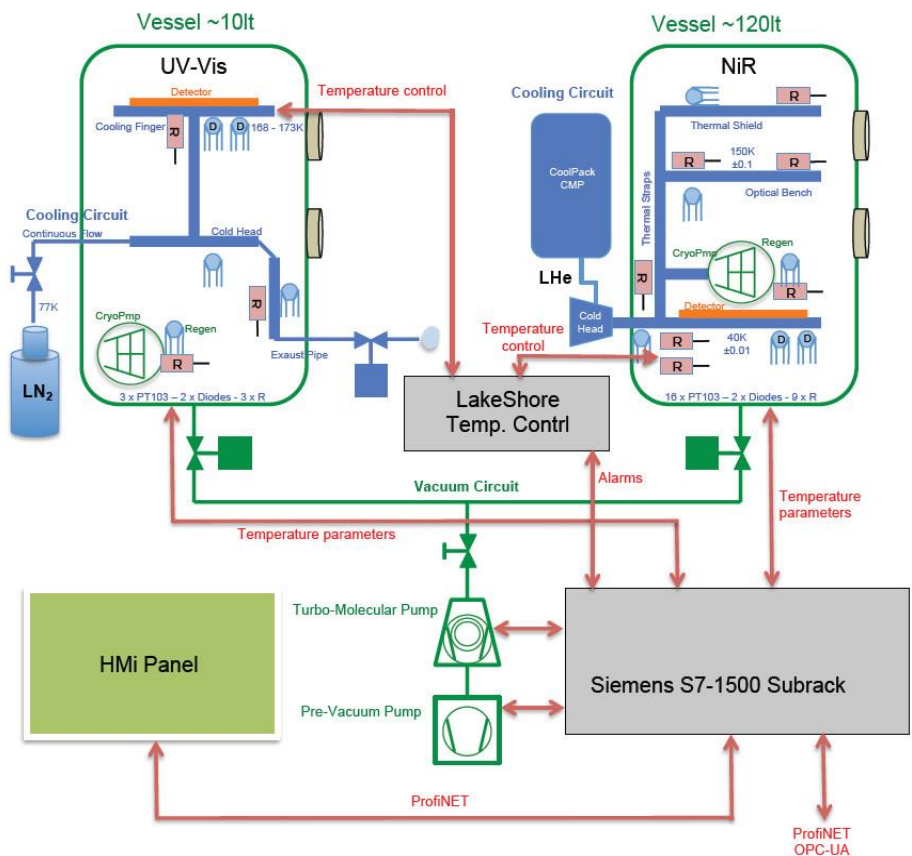

Figure 6. The physical architecture of the SOXS cryo-vacuum control system. Siemens S7-1500 sub-rack is detailed in Figure 7

The Siemens S7-1500 sub-rack box shown in Figure 6 contains all the electronics necessary to control the cryo-vacuum system. The Siemens sub-rack will be inserted in the SOXS electronics cabinets [6]. Figure 7 shows the details of PLC architecture: the PLC-CPU, the power supply, and all the digital and analog I/O to control the system (see also Figure 8). Eight custom electronics boards act as an interface between the PLC and the hardware. Each board is connected to a backplane that to the PLC modules. The boards interface the PLC with the instrumentation and improve its monitoring. Appropriate drivers control the power circuits for heating resistors (Figure 8). The Siemens PLC is then directly connected through Profibus to the turbomolecular pump and to the pressure gauges controller.

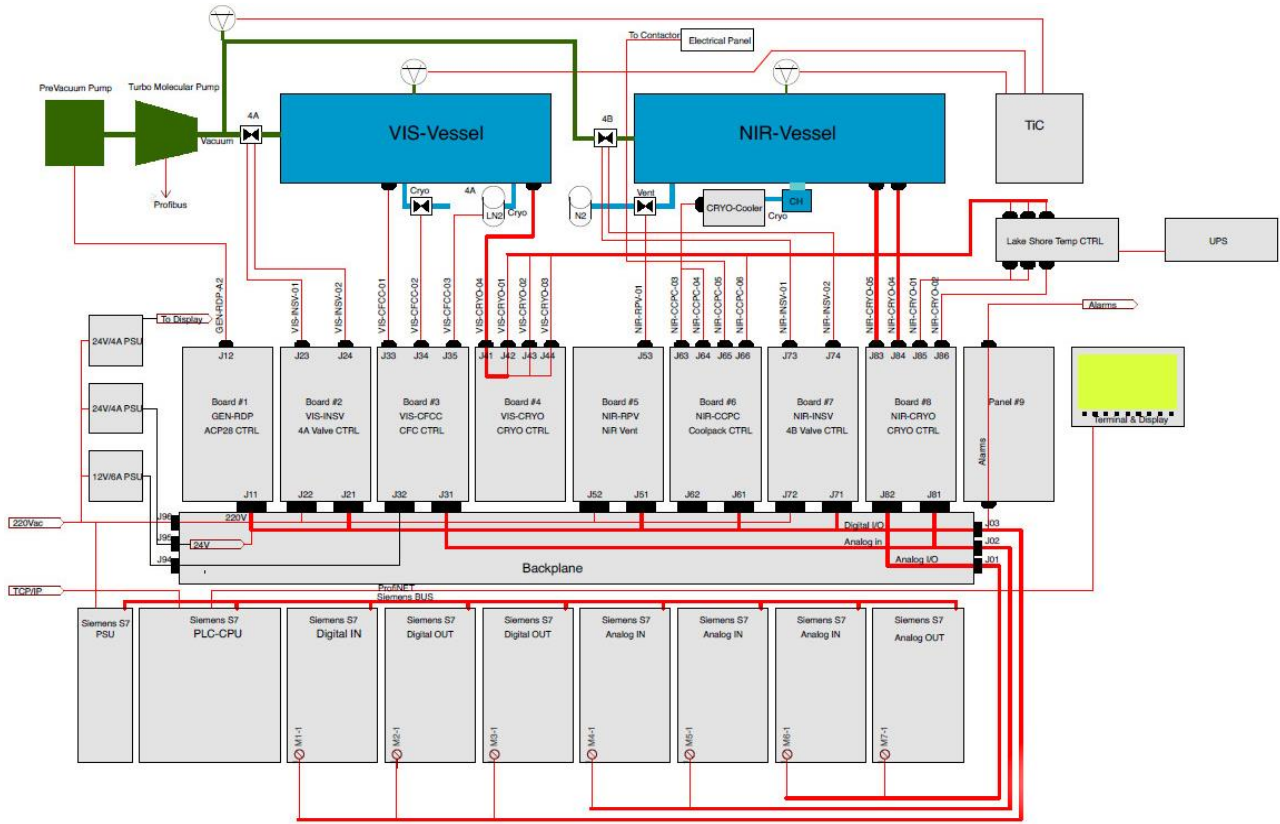

Figure 7. The SOXS vacuum control system architecture. Scheme of control electronics showing the Siemens PLC CPU, PSU and digital and analog I/O and the interface electronics board to connect the PLC to the cryo-vacuum system hardware.

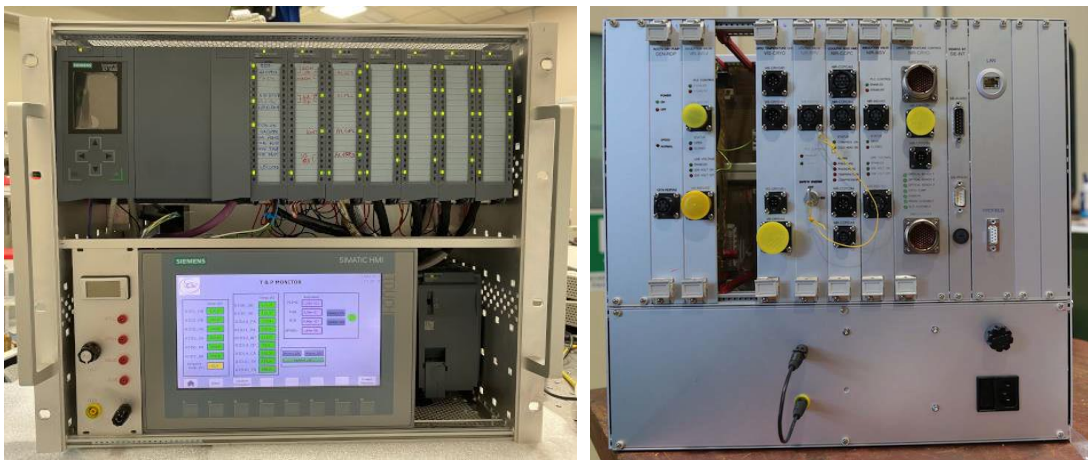

Figure 8. The Siemens sub-rack. Left: front view showing the PLC CPU, PDU, modules and touch panel. Right rear view showing the interface boards to the instrument. The board missing is the CFC one.

### 4.1 Operations of the cryo-vacuum system

The following operations shall be performed by the cryo-vacuum control system through appropriate procedures:

- Evacuation of one vessel or both vessels
- Cooling down of the CCD detector and/or of the NIR cryostat.
- Maintain the CCD and the NIR cryostat operating temperature
- Regeneration of the sorption pumps
- Re-pressurization of one vessel
- Warm up the CCD detector and/or the NIR cryostat.

In the following sections we will describe the procedure to evacuate and cool down the NIR spectrograph and also the detectors thermal control

### 4.2 Evacuation and cooling down of the NIR cryostat

We assume that the NIR cryostat is warm and at atmospheric pressure. In this case the valve (4) in Figure 1 is opened together with the NIR insulation valve (4b) and the pre-vacuum pump is started. When the pressure drops below $10^{-1}$ mbar the turbomolecular pump is started. Then the control system will start the cryocooler only when the pressure is stable below the threshold value of $10^{-4}$ mbar. The cooling of the optical bench and of all the components connected to it is done at the maximum cooling rate possible by the different thermal link without any active control. The only element whose cooling rate is limited is the detector (see next section). The vacuum system continues to support the cryogenic system until the pressure goes below $5 \times 10^{-6}$ mbar and the cryo-pump temperature is below 150 K. When both conditions are met the insulation valve (4b) closes and the vacuum system stops. When the NIR vessel reaches its nominal operating temperature then the cryogenic system starts the thermal control of the optical bench and of the optical elements.

### 4.3 Detectors thermal control

As explained before the detectors thermal control is not in charge of the PLC but of the Lakeshore temperature controller. The thermal control of the CCD UV-VIS detector does not need any special control operation because the CCD detector has not strict limits on the cooling rate.

The same is not true for the NIR detector for which the maximum cooling and warm up rate is set to 2 K/min. To achieve this rate, we act on two fronts: on the one hand we tried to optimize the thermal link between the CCC cold head and the detector housing to have the lowest passive cooling rate compatible with the detector operating temperature, on the other hand we used the Lakeshore ramp function and the heating resistor on the detector cold finger to limit the cooling rate to the desired value. We fixed this value to 0.5 K/min to reduce the time interval during which the detector is cold while the rest of the cryostat is still cooling down.

## 5. THE HUMAN MACHINE INTERFACE

The management of all the cryo-vacuum procedures by the operator will be done through the Simatic HMI touch panel.

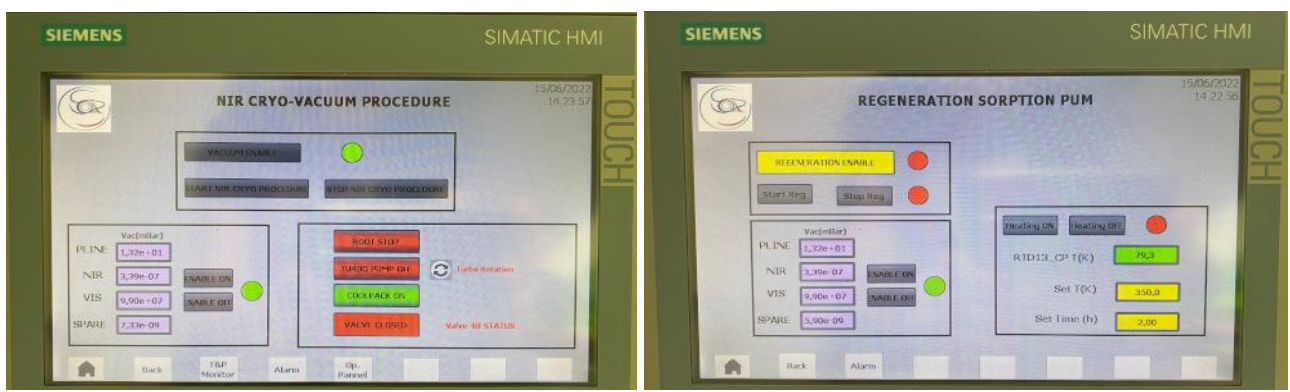

Figure 9. Examples of operator panels. Left: the panel for managing the NIR cryo-vacuum procedure. Right: the panel for managing the regeneration of the cryo-pump.

Most of the panels are devoted to managing the various procedures (Figure 9). Then there are to monitor sensor values or alarms (Figure 10) and, finally, we have synoptic panels that give the status of the system at glance (Figure 11).

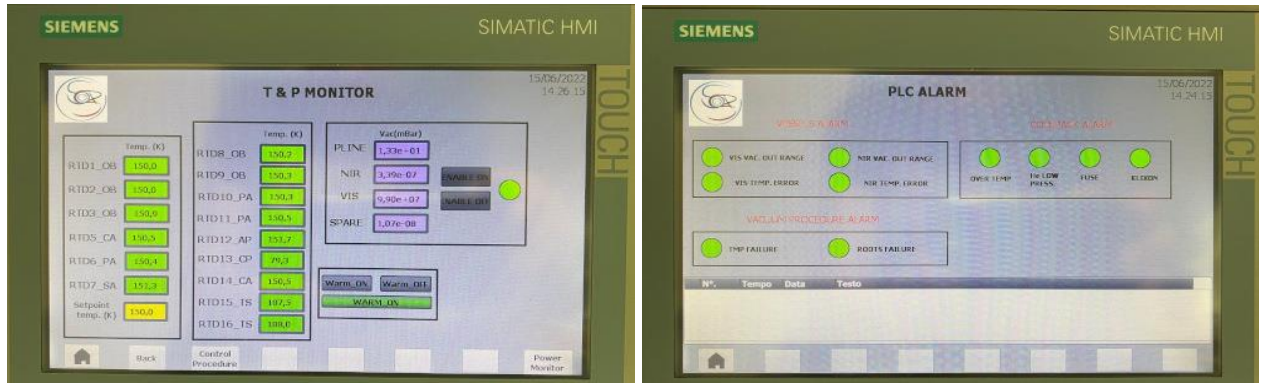

Figure 10. Left: a panel to monitor the temperature and the pressure inside the NIR cryostat and also set the optical bench operating temperature. Right: the alarm panel to show alarms related to the detectors' temperature, the status of the vacuum, that of the cryocooler compressor and of the pumps.

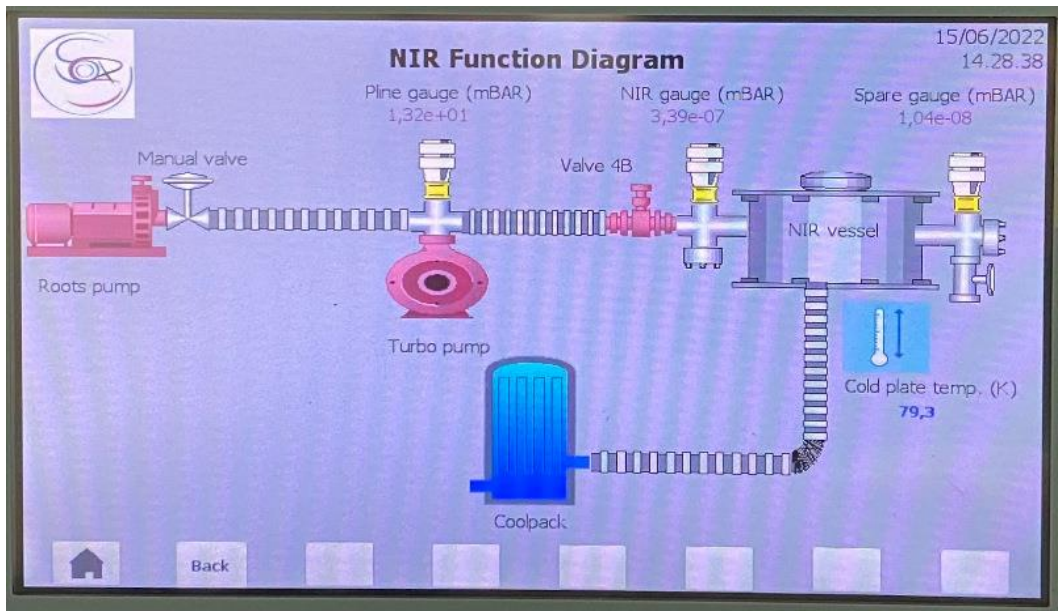

Figure 11. Synoptic view of the status of the NIR Cryo-vacuum system with information on the status of the various devices (pumps, valves, cryocooler compressor) and also the values read by the different sensors (pressure and temperature).

## 6. TESTS OF THE CRYO-VACUUM SYSTEM

The SOXS cryo-vacuum control system is currently undergoing tests at integration site of the NIR spectrograph at the Merate Observatory. So only the functionalities and the preliminary performances related to the NIR cryostat have been checked. The NIR spectrograph is not in its final configuration so the optics, the mechanism and the detector are not present inside the cryostat and have been replaced by mechanical dummies.

### 6.1 Evacuation time and cooling time

Figure 12 shows the pressure evolution versus time inside the NIR cryostat for a typical run. The time necessary to reach the pressure (~$10^{-4}$ mbar) to start to operate the cryocooler is of the order of one hour.

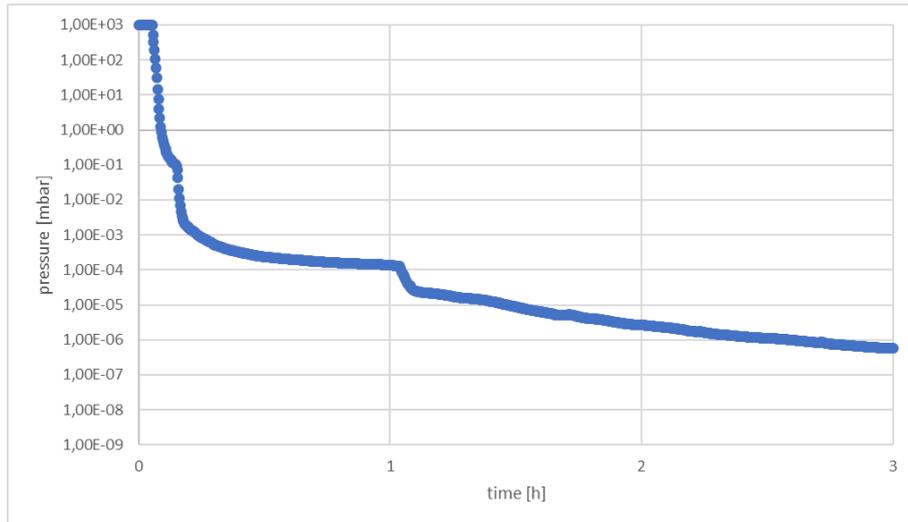

Figure 12. Evacuation time of the NIR cryostat. The steep decrease in pressure around $10^{-1}$ mbar is due to the switching on of the turbo-molecular pump, that at $10^{-4}$ mbar is due instead to the switching on of the cryocooler.

Figure 13 shows the evolution of the temperature in the NIR cryostat during the cooling down procedure as measured by the sensors on the thermal shield, the optical bench and the cryo-pump.

Due the current conductivity of the thermal links between the cryocooler head and the NIR cryostat the time necessary to reach the operating temperature is almost four days, well above the requirement of 30 hours.

An increase in the conductivity of the thermal links on the optical bench is under consideration not only to decrease the cooling time but also to make the cooling rate of the detector and that of the rest of the cryostat more similar to each other (see also section 4.3).

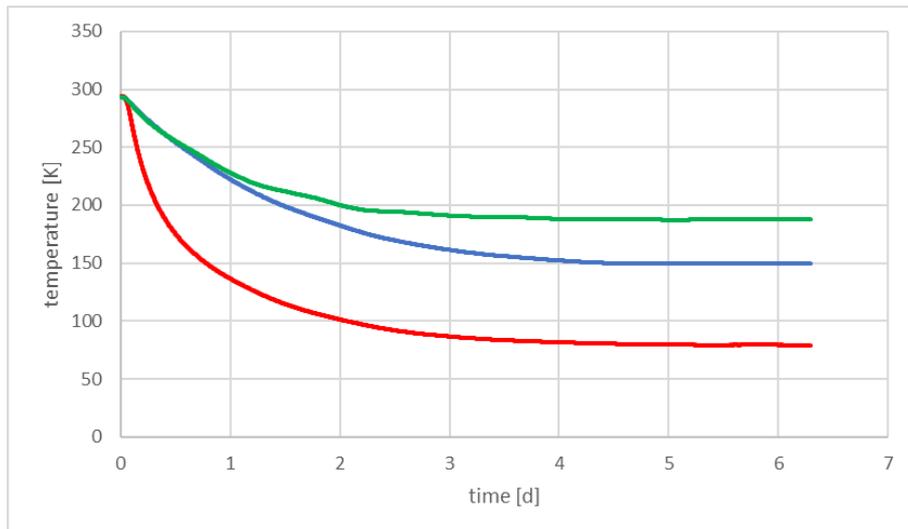

Figure 13. Cooling time of the NIR cryostat. Red line is the temperature of the cryo-pump, blue line is the temperature of the optical bench and the green line is the temperature of the thermal shield

### 6.2 Vacuum stability

Vacuum stability has been measured only for intervals of few days. Figure 14 shows the evolution of the pressure measured inside the NIR cryostat during a six days period. No degradation of the vacuum is visible during this period of time. A periodic trend with varying amplitude seems to be present and is maybe related to day-night cycle.

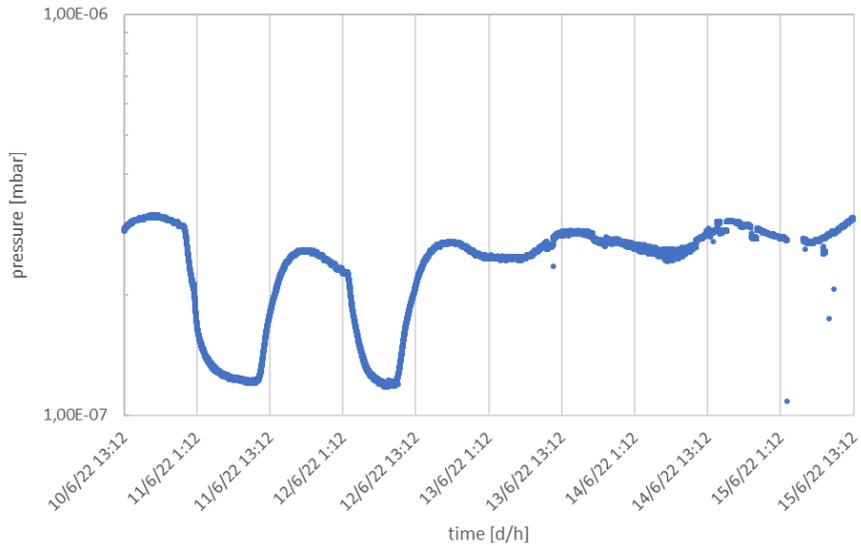

Figure 14. Evolution of the pressure inside the NIR cryostat during a 6 days period.

### 6.3 Temperature stability of the NIR cryostat

The long cooling time of the NIR cryostat, discussed in section 6.1, brings as side effect a good temperature stability with time. In fact, the cryostat operating temperature is maintained with little or no intervention of the active thermal control system with very small power dissipated on the heating resistors. Figure 15, on the left panel, shows the evolution of the temperature on the NIR cryostat optical bench and on two optical assemblies, the camera assembly and the prisms assembly). The readings of the temperature sensors use the standard calibration curve provided by the Lakeshore for PT-100 sensors, so the differences in temperature can be partly due to calibration issues. The plot shows that the temperature of the bench and that of the camera assembly is well within the 150 ± 2 K requirement. The temperature of the prisms' assembly is between 150.3 K and 150.4 K so even if the average value is above 150 K, its variation is well within the ±0.1 K requirement.

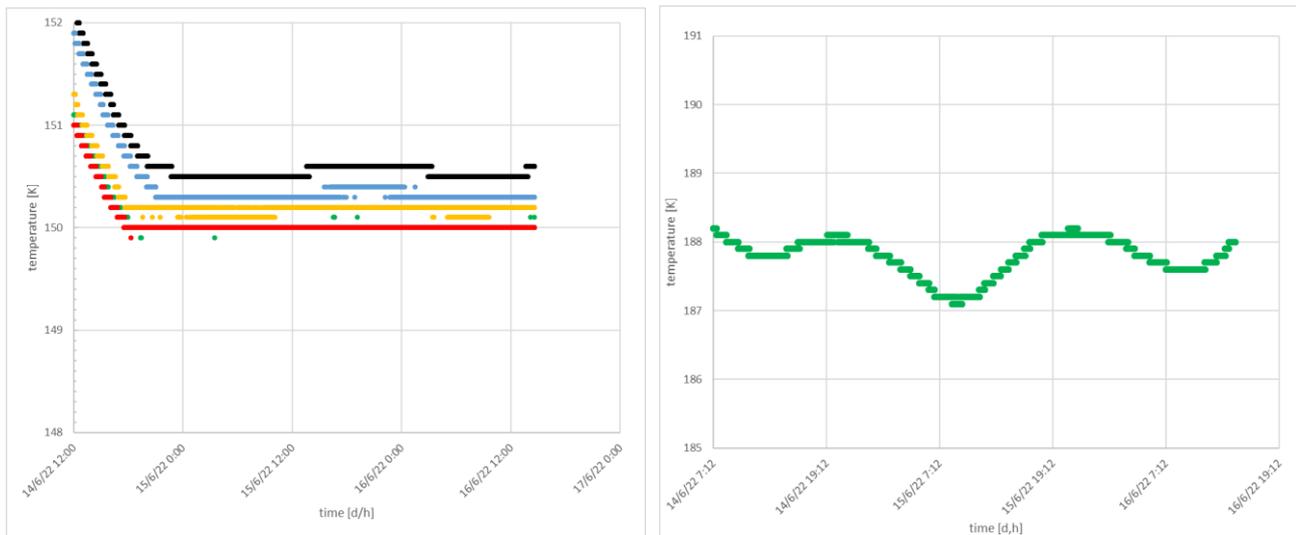

Figure 15. Left panel: evolution of temperature of the optical bench of the NIR cryostat. Red, green and yellow dots refer to three of the five sensors distributed along the optical bench. Blue dots refer to the temperature on the prisms' assembly and black dots to that of the camera assembly. Right panel: the evolution of the temperature on the thermal shield of the NIR cryostat. No resistor is present on the thermal shield.

The right panel of Figure 15 shows the temperature of the thermal shield. The thermal shield is left floating as no heating resistors are mounted on it.

### 6.4 Regeneration of NIR vessel cryo-pump

Figure 16 shows the variations of the cryostat internal pressure and of the cryo-pump temperature during the regeneration process. The set value for the temperature was 350 K while the time interval for this test was 12 hours.

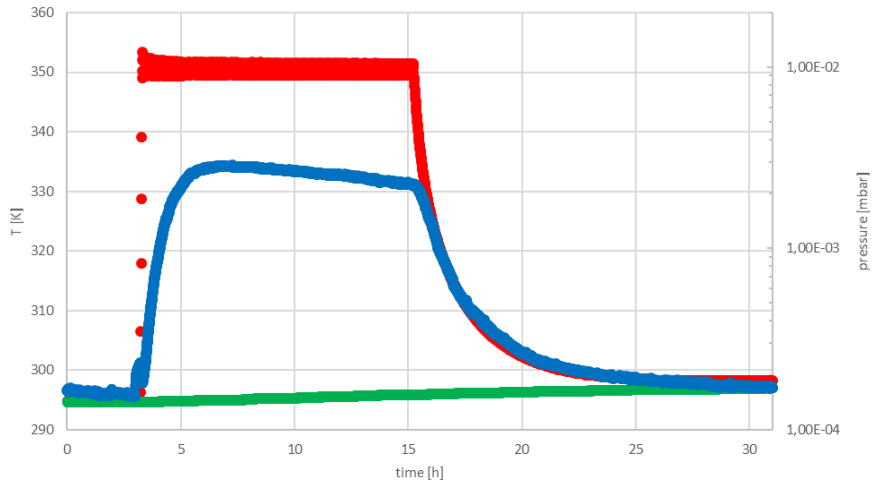

Figure 16. Regeneration of the cryo-pump. Blue dots show the temperature of the cryo-pump. Green dots are the values of the pressure inside the vessel. Orange dots shows the increase in temperature of the detector during the regeneration procedure.

The reason to have such a long period for the process was to investigate the behavior of the detector's temperature. In fact, in the plot it is also shown the temperature of the detector as there was some concern on the possibility that the detector could heat up during the procedure as it is thermally linked to the cryo-pump through the CCC cold head. The plot clearly shows that the increase in temperature of the detector is negligible.